\documentclass[preprint,floatfix]{revtex4}
\usepackage{graphicx}
\usepackage{dcolumn}
\usepackage{bm}
\usepackage{amsmath,amssymb}

\newcommand{\beqn}{\begin{equation}}
\newcommand{\eeqn}{\end{equation}}

\newcommand{\flow}{s}

\begin{document}
\title{The impact of bound states on\\ similarity renormalization group transformations}
\author{ Stanis{\l}aw D. G{\l}azek } \email{stglazek@fuw.edu.pl}
\affiliation{Institute of Theoretical Physics, University of Warsaw, Poland}
\author{ Robert J. Perry } \email{perry.6@osu.edu}  
\affiliation{Department of Physics, The Ohio State University, Columbus, OH 43210}
\preprint{IFT/04/08}
\date{Draft of March 19, 2008 }

\begin{abstract} We study a simple class of
unitary renormalization group (RG) transformations
governed by a parameter $f$ in the range $[0,1]$.
For $f=0$, the transformation is one introduced by
Wegner in condensed matter physics, and for $f=1$
it is a simpler transformation that is being used
in nuclear theory. The transformation with $f=0$
diagonalizes the Hamiltonian but in the
transformations with $f$ near 1 divergent
couplings arise as bound state thresholds emerge.
To illustrate and diagnose this behavior, we
numerically study Hamiltonian flows in two simple
models with bound states: one with asymptotic
freedom and a related one with a limit cycle. The
$f=0$ transformation places bound-state
eigenvalues on the diagonal at their natural
scale, after which the bound states decouple from
the dynamics at much smaller momentum scales. At
the other extreme, the $f=1$ transformation tries
to move bound-state eigenvalues to the part of the
diagonal corresponding to the lowest momentum
scales available and inevitably diverges when this
scale is taken to zero. Intermediate values of $f$
cause intermediate shifts of bound state
eigenvalues down the diagonal and produce
increasingly large coupling constants to do this.
In discrete models, there is a critical value,
$f_c$, below which bound state eigenvalues appear
at their natural scale and the entire flow to the
diagonal is well-behaved. We analyze the shift
mechanism analytically in a 3x3 matrix model,
which displays the essense of this RG behavior and
we compute $f_c$ for this model. 
\end{abstract}
\maketitle

\section{ Introduction }
\label{sec:intro}

Wilsonian renormalization group transformations
typically eliminate (integrate out) degrees of freedom
whose energy is much higher than those of interest,
replacing them with effective scale-dependent interactions.
Such transformations allow one to tune an effective theory
resolution, focusing on essential degrees of
freedom and interactions at any scale of interest.
Anticipated by Kadanoff's block spin
transformation~\cite{Kadanoff:1965}, early
transformations explicitly reduced the number of
degrees of freedom by lowering cutoffs on
energy~\cite{Wilson:1970,Wilson:1975}. G{\l}azek
and Wilson introduced a similarity renormalization
group (SRG) procedure which instead uses
transformations that do not remove any degrees of
freedom but eliminate couplings between disparate
energy scales~\cite{Glazek:1993,Glazek:1994,Glazek:1998}.
After regularization and identification of necessary counterterms,
the SRG procedure eventually produces a renormalized,
band-diagonal matrix representation of the
Hamiltonian. Independently, Wegner introduced
non-perturbative differential flow equations that
unitarily transform Hamiltonian matrices to the
band-diagonal form \cite{Wegner:1994,
Kehrein:2006}, and Wegner's flow equations can be
employed in the SRG procedure. 

A simplified version of Wegner's
transformation~\cite{PerrySzpigel} has been
successfully applied to a number of nuclear
few-body
problems~\cite{Bogner:2006srg,Bogner:2007jb}.
Nuclear many-body calculations are plagued by
strong nucleon-nucleon correlations due to a hard
core and strong short-range tensor force, so
perturbative and variational methods converge
poorly. The new simple transformation produces
universal nucleon-nucleon interactions with
drastically improved perturbative and variational
behavior. It has also been shown that one can apply SRG with
Wegner's flow (or a suitably altered flow) for studying
the connection between asymptotic freedom and limit
cycles~\cite{lcoet} and it has been suggested that an infrared 
limit cycle may exist in QCD~\cite{Braaten}. These examples indicate 
that convergence properties of SRG transformations are 
relevant to the theory of particles over a broad range of 
energies.

We find that convergence properties of the simple
transformation are worse than those of Wegner's
transformation. The simple transformation tends to
diverge whenever the SRG parameter approaches the
momentum scale at which a bound state is formed in
the theory. This effect may present no problem for
applications in low-energy nuclear physics as long
as the SRG parameter stays larger than
the momentum scales at which formation of bound
states occurs. However, such fortunate conditions
may not be available if the Efimov
effect~\cite{Efi70,Efi71,Thomas} in the three-nucleon
problem~\cite{BHK99,BHK99b,BHK00,Mohr:2006} cannot
be avoided. The Efimov effect shows up as a limit
cycle. We will see that Wegner's transformation is
capable of resolving limit cycle behavior, while
the simple transformation diverges as the first
high-energy cycle is resolved.

Transform a Hamiltonian $H(0)$ using a unitary 
operator $U(\flow)$,
\beqn
   H(\flow) = U(\flow) H(0) U^\dagger(\flow) \, ,
\eeqn
where $\flow$ is the SRG flow parameter. We want to 
choose $U(\flow)$ so that $H$ is diagonalized as $\flow 
\rightarrow \infty$
(band-diagonalized for finite $s$). We choose $s=0$ for the initial 
value, so $H(0)$ can be thought of as an input bare regularized 
Hamiltonian with all required and properly adjusted 
counterterms (such Hamiltonians are established 
using the same SRG procedure but this aspect is not 
in our focus here). 

Taking derivatives of both sides of Eq. (1), we see 
that $H(s)$ evolves according to
\beqn
  \frac{dH(\flow)}{d\flow}
    = [\eta(\flow),H(\flow)] \ ,
\eeqn
with
\beqn
   \eta(\flow) = \frac{dU(\flow)}{d\flow} U^\dagger(\flow) 
          = -\eta^\dagger(\flow)
   \ .
\eeqn
Choosing $\eta(\flow)$ specifies the transformation. 
We study transformations that mix two simple choices,
\beqn
\label{G=D}
  \eta(\flow) =  [D(\flow), H(\flow)] \, ,
\eeqn
where $D(\flow)$ is the diagonal part of $H(\flow)$ 
in momentum representation, and
\beqn
\label{G=T}
  \eta(\flow) =  [T, H(\flow)] \, ,
\eeqn
where $T$ is a fixed matrix, here chosen to be the 
kinetic energy.
If one considers $T$ to be an arbitrary $H_0$ 
that has a known spectrum, the interaction is the 
remaining part, $H_I = H-H_0$. 

Using $D$ in $\eta$ was Wegner's initial choice
and this transformation has been studied
extensively~\cite{Kehrein:2006}. The use of $T$
was explored perturbatively~\cite{PerrySzpigel}
and then shown to effectively decouple low- and
high-momentum scales in a universal
characterization of the nucleon-nucleon
interaction~\cite{Bogner:2006srg,Bogner:2007jb}.
Universality in effective nucleon-nucleon
interactions was discovered earlier~\cite{VlowkRG,Vlowk2} 
using the same transformation Wilson used in his initial
numerical RG calculations~\cite{Wilson:1970}. 

We will see that using $T$ instead of $D$ produces
singularities starting at bound state thresholds
and limits how far the transformation can be run.
But we stress that if it is not run too far, for
any phenomenologically tuned nucleon-nucleon
interaction $H(0)$ the low-energy part of
$H(\flow)$ is nearly universal. This means that
all infrared-constrained nucleon-nucleon
interactions with bare cutoffs at or even well
above 500 MeV collapse onto a nearly universal
infrared Hamiltonian after the SRG parameter,
$\lambda=1/\sqrt{s}$,
playing the role of effective cutoff, is evolved
to well below the bare cutoff. These evolved
potentials disagree only on the high-energy part
of $H(\flow)$ , which is not constrained by
low-energy physics~\cite{Bogner:2007jb}.

Introducing a parameter $f$ which takes values in 
the range $[0,1]$, we define
\beqn
\label{G}
G_f(\flow) = f T + (1-f) D_f(\flow),
\eeqn
and write
\begin{eqnarray}
\label{SRG}
\frac{d}{ds} \, H_f(s) & = & [F\{H_f(s)\},H_f(s)] \, , \\
H_f(0) & = & H \, ,
\end{eqnarray}
where $H$ is an initial Hamiltonian matrix and the 
SRG generator matrix takes the form
\begin{eqnarray}
F\{H_f(s)\} & = & [G_f(s), H_f(s)] \, , 
\end{eqnarray}
and thus defines $\eta$ that depends on $f$ and 
interpolates between the two cases from Eqs.~(\ref{G=D}) 
and (\ref{G=T}) when $f$ varies from 0 to 1.  
For explicit calculations one uses a basis in 
which $T$ is diagonal since it is in this 
representation that $H(\flow)$ is driven towards 
band-diagonal form. 

The generator $F\{H_f(s)\}$ is guaranteed to bring the
Hamiltonian to diagonal if some sufficient and
easily verifiable conditions are satisfied. Such
conditions will be discussed after we introduce
all details that count in the derivation, in the context of two
examples of basic interest in physics.

Namely, we use the family of generators $F\{H_f(s)\}$ to
evolve Hamiltonian matrices that exhibit
asymptotic freedom and limit cycle behavior. The
asymptotically free matrix model is easily derived
from the nonrelativstic Schr{\"o}dinger equation
in two dimensions with a delta-function potential.
Isolate the angular momentum zero states and
discretize the momentum so that $p \rightarrow b^n
p_0$ with $b>1$; include the appropriate weights
to reproduce the momentum representation bound-state 
integral equation in the limits where $n$ is
allowed to be any integer and $b \rightarrow
1$. Introduce cutoffs so that $M \le n \le N$ and
you have the matrix we use to illustrate
asymptotic freedom ($M$ is a large negative and
$N$ is a large positive integer number). The limit
cycle model is obtained by adding an imaginary
part to the same asymptotically free model
Hamiltonian. The operator $T$ can be replaced by
any Hermitian operator one wants to use and the
methods we employ in this study can still be applied.

\section{ Details of equations }
\label{details}

The equations for a fixed value of $f$ contain the
diagonal matrix $G_f(s)$ given in Eq. (\ref{G})
that contains the diagonal part of $H_f(s)$. The
Hamiltonian matrix is split into its diagonal and
off-diagonal parts at every value of $s$, 
\begin{eqnarray}
H_f(s) & = & D_f(s) + V_f(s) \, .
\end{eqnarray}
This splitting implies also 
\begin{eqnarray}
H_f(s) & = & G_f(s) + [H_f(s) - G_f(s)] \\
& = & G_f(s) + f[D_f(s)-T] + V_f(s) \, .
\end{eqnarray}
This means  that the diagonal part of the interaction 
is included in $D_f(s)$. The important point is 
that only $V_f(s)$ has non-zero off-diagonal matrix 
elements. Diagonal matrix elements of $V_f(s)$ are 
zero. $T$ and $D_f(s)$ are diagonal and we use an 
abbreviated notation for their matrix elements: 
$T_{mn} = T_m \delta_{mn}$, $D_{f mn}(s) = D_m \delta_{mn}$, 
and $G_{f mn}(s) = G_m \delta_{mn}$. Our abbreviated notation for 
interaction matrix elements is $V_{f mn}(s) = 
(1-\delta_{mn})V_{mn}$, where $V$ is the full 
interaction part in the matrix $H_f(s) = T + V$.

The SRG Eq. (\ref{SRG}) implies 
\begin{eqnarray}
\label{SRGD}
\frac{d}{ds} \, D_n & = & 2 \sum_k (G_n - G_k) V_{nk} V_{kn} \, , \\
\label{SRGV}
\frac{d}{ds} \, V_{m \neq n} & = & - (G_m - G_n)(D_m - D_n) V_{mn} + \sum_{m \neq k \neq n} (G_m + G_n - 2G_k) V_{mk} V_{kn} \, .
\end{eqnarray}
We solve these equations numerically, starting from 
\begin{eqnarray}
\label{Hinitial}
H_f(0)_{mn} & = & \sqrt{ E_m E_n} \, 
\left[ \, \delta_{mn} - g - i h \, {\rm sgn}(m-n) \, \right] \, , \\
E_n & = & b^n \, , \\
M \le &n& \le N \, . 
\end{eqnarray}

In the case of the model with asymptotic freedom, 
the coupling constant $h$ is set to 0, the Hamiltonian 
matrix is real and Eqs. (\ref{SRGD}) and (\ref{SRGV}) 
display all relevant formulae. In the case of the model 
with a limit cycle, the Hamiltonian contains an imaginary 
part and the equations we use to compute the Hamiltonian 
flow numerically involve the imaginary components,
\begin{eqnarray}
H_f(s)_{mn} & = & r_{mn} + i c_{mn} \, , 
\end{eqnarray}
where $r$ is a real symmetric matrix and $c$ is a real antisymmetric 
matrix. Thus, $D_m = r_{mm}$. With this notation, the SRG equations we solve are
\begin{eqnarray}
\label{SRGrnn}
\frac{d}{ds} \, D_n & = & 2 \sum_k (G_n - G_k) (r_{nk}^2 + c_{nk}^2) \, , \\
\label{SRGrmn}
\frac{d}{ds} \, r_{m\neq n} & = & - (G_m - G_n)(D_m - D_n) r_{mn} + \sum_{m\neq k \neq n}(G_m + G_n - 2G_k) (r_{mk} r_{nk} + c_{nk} c_{mk}) \, , \nonumber \\
& & \\
\label{SRGcmn}
\frac{d}{ds} \, c_{mn} & = & - (G_m - G_n)(D_m - D_n) c_{mn} + \sum_{m\neq k \neq n} (G_m + G_n - 2G_k) (c_{mk} r_{nk} -  c_{nk} r_{mk} ) \, . \nonumber \\
& & 
\end{eqnarray}

We will show that to a good approximation $H_f(s)$
retains the form of $H_f(0)$ for matrix elements
between states with kinetic energies well below
the SRG parameter $\lambda = 1/\sqrt{s}$. The
dominant change that occurs in these matrix
elements is that $g$ is replaced by $g_f(s)$. This
feature emerges in numerical calculations. We also
analytically derive this result in the limit of
large $b$ in the next section. 

Once it is established through numerically calculated 
non-perturbative evolution of the whole Hamiltonian 
that its evolution can be reduced to the evolution of 
the coupling constant $g_f(s)$, we focus discussion on 
the evolution of the coupling and its dependence on $f$. 
The coupling constant is defined as
\begin{eqnarray}
g_f(s) & = & 1 - H_f(s)_{MM}/E_M \, ,
\end{eqnarray}
where $E_M$ is the smallest energy in the matrix
representation of the theory (smallest allowed
eigenvalue of $T$) and $H_f(s)_{MM}$ is the
smallest energy diagonal matrix element of
$H_f(s)$, the infrared corner of the Hamiltonian.
We can choose any diagonal or off-diagonal matrix
element near the infrared corner of the Hamiltonian
matrix to define the same coupling constant and no
significant changes result until this element
begins to freeze at its asymptotic limit as $s \rightarrow \infty$. Each
matrix element follows a universal trajectory
until it freezes at its $s \rightarrow \infty$
value. Only the lowest diagonal element displays the
full evolution of the diagonal.

When $G$ differs from $D$, i.e., $f > 0$, we
show that the SRG transformation does not
necessarily bring the Hamiltonian matrix to
diagonal form as $s \rightarrow \infty$. 
However, there is a sufficient condition for 
cases with $f>0$ that will be derived in 
Section~\ref{sec:3x3}. Namely,
\begin{eqnarray}
\frac{ d V}{ d T} > -1 \, ,
\end{eqnarray}
where the derivative means the rate of change of 
matrix elements of $V$ along the diagonal in units 
of rate of change of eigenvalues of $T$ along the 
diagonal. This condition could be violated when 
bound-state (negative) eigenvalues appear on the 
diagonal among positive eigenvalues. To study how 
the SRG transformation behaves depending on the 
choice of $f$ in the presenece of bound states, 
and in particular what happens in the case $f=1$ 
that is useful in nuclear physics, one needs to 
see what happens in the generic models when $f$ 
deviates from 0 and reaches 1. Since the sufficient 
condition could be violated when the SRG parameter 
$\lambda$ passes the momentum scale of binding, one should 
find out how $H_f(s)$ behaves around $s$ corresponding 
to this region.

In the next section we discuss results of
calculations of $H_f(s)$. We show that Wegner's
transformation ($f=0$) encounters no difficulty
and places bound-state eigenvalues on the diagonal
when $\lambda$ approaches the appropriate
bound-state scale from above. The $f=1$
transformation breaks down as the effective cutoff
approaches a bound-state scale. In fact, the
transformation moves the bound-state
eigenvalue to the infrared corner of the
Hamiltonian. When the minimal momentum scale is
much less than the bound-state momentum scale, the
$f=1$ transformation forces
off-diagonal matrix elements to diverge in order to
move the bound-state eigenvalue to such a significantly {\it
wrong} scale. 

For values of $f$ between these two extremes the
transformation puts the bound-state eigenvalue on
the diagonal at some scale between the natural one
and the infrared cutoff. However, for $f$
approaching 1, the amount of shift approaches the
maximal possible and correspondingly large couplings 
must be generated on the way. For $f=1$, the 
transformation becomes numerically unstable near 
the bound-state scale when the infrared cutoff 
tends to zero. 

In discrete matrix notation, one can ask how large 
a value of $f>0$ causes the first shift of the bound-state 
eigenvalue down the diagonal, just by one free 
energy level in comparison to the Wegner case ($f=0$). 
This value of $f$ will be called critical, and denoted 
by $f_c$. It can be computed numerically to high 
accuracy. 

\section{ Numerical results: asymptotic freedom}
\label{sec:results}

The characteristic SRG behavior displayed by the
full matrix $H_f(s)$ is difficult to grasp (in our
typical cases, the matrix has about 2000 matrix
elements, each a function of $s$ that ranges from
0 to $\infty$) without studying a number of cases.
Such studies involve large amounts of data.
Fortunately, the overall result of such studies is
that the essence of what happens for large
matrices can be explained in a simple way using
just one running coupling constant. Most
interestingly, the characteristic behavior of this
one running coupling constant in large matrices
can be explained using much smaller matrices, and
a matrix that is only 3x3 in size will be
sufficient. These two facts guide the way we
present and discuss our results.

Our goal is to describe the most important qualitative 
features of the Hamiltonian evolving from its initial 
form of the type $\sqrt{E_m E_n}~(\delta_{mn}-g)$ to 
complete diagonalization. We need to cover a large 
range of scales and make important features of $V$ 
at all scales visible simultaneously. Taking our cue from 
the initial Hamiltonian and experience gathered in 
observing many examples of the SRG flows, we display 
what we will call the {\it scaled interaction}, 
${\cal V}_{mn} = V_{mn}(s)/\sqrt{E_m E_n}$. Matrix elements 
of the scaled interaction are all of ${\cal O}(1)$ or 
decay to 0 during the entire process of diagonalization 
when Wegner's transformation is used ($f=0$). When $f=1$ 
however, matrix elements of $\cal V$ diverge as the 
bound-state threshold is reached (see below).

Throughout its evolution, the scaled interaction
matrix can be approximated by $ {\cal
V}_{mn} \sim - g_f(s)+ corrections$ for subscripts
$m$ and $n$ below the transition region, in which
the subscripts $m$ and $n$ take values $k$ such
that $sE_k^2 \sim 1$. Well above the transition
region the Hamiltonian matrix is diagonalized. As
$s$ increases, $g_f(\flow)$ increases to a maximum, at
which point the bound-state negative eigenvalue
emerges on the diagonal or a process of shifting
the bound-state eigenvalue down the diagonal
begins.

We start with the model case of asymptotic freedom 
(a discretized $\delta$-function in two dimensions), 
for which $h$ in eq. (\ref{Hinitial}) is zero and 
the single bound-state eigenvalue determines $g$ 
through dimensional transmutation. In the notation of
Section \ref{details}, $c_{mn}=0$ and $r_{mn}(s=0)
=\sqrt{E_m E_n}(\delta_{mn}-g)$. We consider
$g=0.0400022797581725654$. (This particular value
of $g$ is not significant; it is found from the
condition that for $h=0$, $b=4$, $N=16$, $M=-25$,
the bound-state eigenvalue is with high accuracy
the same as one of the bound-state eigenvalues in
another case: one with a limit cycle for $g=0$ and
$h = \tan{\pi/50}$, and eigenvalue $E \sim
-7.644479 \, 10^{-6}$, see \cite{lcoet}.) 

\begin{figure}[ht]
\includegraphics[scale=0.5]{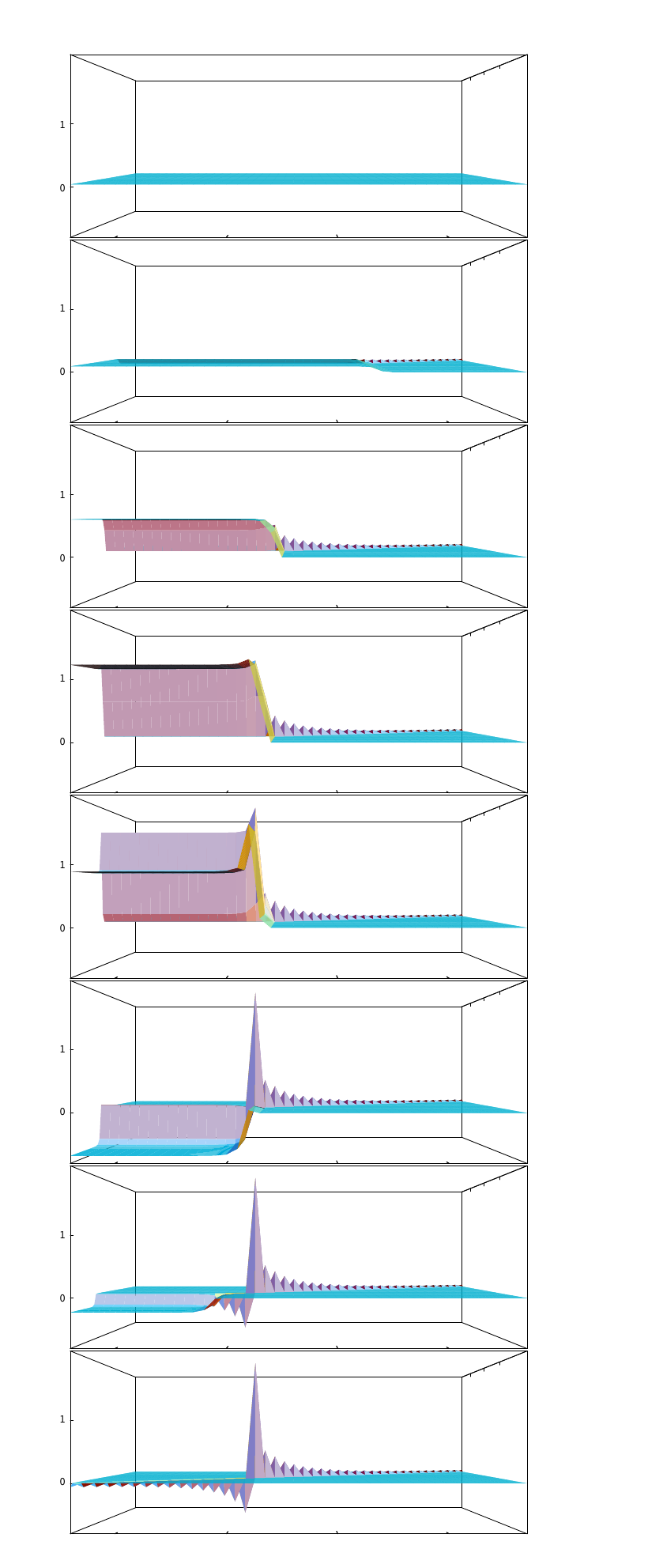}
\caption{\label{fig:fullh0} SRG evolution of 
$H=T+V$ with $\lambda$ for $f=0$. Successive 
frames correspond to entries in Table \ref{tab:gl}. 
The horizontal axes display subscripts $i$ and $j$ 
(running from $M$ to $N$, counting from left to 
right and from front to back) of the matrix $-V$ 
and the vertical axis displays corresponding matrix 
elements $-V_{ij}$ divided by $\sqrt{E_i E_j}$. 
See text for details. }
\end{figure}
In Fig. \ref{fig:fullh0}, we display frames from a 
movie of the evolving interaction, $V(s)$, in the 
Hamiltonian $H=T+V$, using Wegner's transformation, 
$f=0$. We show $-{\cal V}_{mn}$. Since 
small-energy matrix elements of $\cal V$ equal the negative of the coupling 
constant, we need to display $-\cal V$ instead of 
$\cal V$ itself in order to directly show how well 
the one coupling constant $g_{f=0}(s)$ approximates 
the evolution of $\cal V$ below the transition region 
and at the same time show how the coupling constant 
itself evolves. With these rescaling and display 
conventions, $T$ is the identity matrix. To get 
$H_{mn}/\sqrt{E_m E_n}$, simply take the negative of 
the displayed values of $-\cal V$ and add peaks of 
height one along the diagonal in each frame.
Details concerning the frames shown in Fig. 
\ref{fig:fullh0} are listed in Table \ref{tab:gl}.
\begin{table}[ht]
\caption{\label{tab:gl} SRG parameters $g$ and 
$\lambda$ for frames shown in Fig. \ref{fig:fullh0}, 
numbered from the top to bottom. In this example, 
$g=0.040002$, $h=0$, $b=4$, $M=-25$, $N=16$, and 
all displayed numbers are rounded to 6 decimal places.}
\begin{ruledtabular}
\begin{tabular}{lll}
frame           &  $\ln(\lambda)/\ln{b}$ & $g(\lambda)$        \\
\hline
1 (top)         &  22.780321    &  0.040002  \\
2               &   2.766096    &  0.092055  \\
3               &  -6.864809    &  0.600768  \\
4               &  -8.369638    &  1.234710  \\
5               &  -8.570281    &  0.891475  \\
6               &  -9.071891    & -0.680443  \\
7               & -12.282193    & -0.226083  \\
8 (bottom)      & -27.330482    & -0.060769  \\
\hline 	
\end{tabular}
\end{ruledtabular}
\end{table}

The negative of the scaled interaction is initially 
a featureless plane (frame 1), its size fixed at 
about 0.04, just above zero because it is a 
negative of a momentum representation of an 
atractive delta-function in position representation 
with the initial value of the coupling constant 
$g \sim 0.04$. As $s$ increases from 0, the plane 
drops to zero at highest energies, creating a 
cliff (frame 2) between the low and high energy 
parts of the matrix (transition region). This cliff 
runs along a single row and column that meet on the 
point along the diagonal at which newly decoupled eigenvalues 
are emerging. The cliff moves towards lower energies 
and grows in height as $s$ increases (frame 3), 
showing evolution of the transition region between 
a flat plane of zeroed high-energy off-diagonal 
matrix elements of ${\cal V}$ and a low-energy 
plateau that is rising higher as $s$ increases 
(that this plateau rises means that the coupling 
constant increases and the potential itself becomes 
more negative). The positive peaks left along the 
high-energy diagonal decouple from the rest of the 
matrix when the ridge moves past and they are 
basically left at the height of the low-energy 
rising plane at the decoupling point.

This process continues, with the low-energy part of 
the matrix rising and  high-energy eigenvalues being 
left in isolation on the diagonal as the growing 
cliff separating low and high energy portions of 
the Hamiltonian moves towards the infrared corner of 
the matrix and the off-diagonal high-energy part of 
the matrix settles to zero. 

We display a frame in which many high-energy eigenvalues 
are in place and the low-energy plateau has barely 
fallen from its maximum height (frame 4), at a point where a bound state is 
going to emerge on the diagonal of $H(s)$. Remember 
that what we are showing is $-{\cal V}$, which must 
be multiplied by $\sqrt{ E_m E_n}$ and subtracted from 
$T$ to produce $H(s)$. Two important things happen: 
the low-energy plane is high enough to cancel $T$ and 
produce a negative eigenvalue on the diagonal, and as 
this bound-state eigenvalue is left on the diagonal 
the low-energy plane reverses its motion (frames 4 
and 5) and drops rapidly to negative values (frame 
6, the coupling constant becomes negative and the 
whole interaction $\sim – g_f(s)$ becomes positive, 
i.e., repulsive).

After the bound-state threshold is crossed, $V$ 
adds to the eigenvalues of $T$, while up to 
this point it had subtracted from these eigenvalues. 
After quickly reaching its deepest level,
the low-energy plane gradually rises to zero (frames 
7 and 8), leaving a sequence of eigenvalues that 
smoothly emerge from this flow one after another 
toward the infrared corner. 

The only violent changes in these scaled variables 
appear around the point where the bound state emerges. 
This is where the coupling that 
characterizes the evolution of the low-energy part of 
the matrix grows to its maximum value, before 
dropping rapidly to negative values.

It is clear that the evolution of the entire matrix $\cal V$ 
is well-described by the evolution of $g_f(s)$ in the 
case of $f=0$. The same is true in all other cases we 
consider. Therefore, we will present only the functions 
$g_f(s)$ in all these cases rather than 
frames from many movies. 

\begin{figure}[ht]
\includegraphics[scale=0.5]{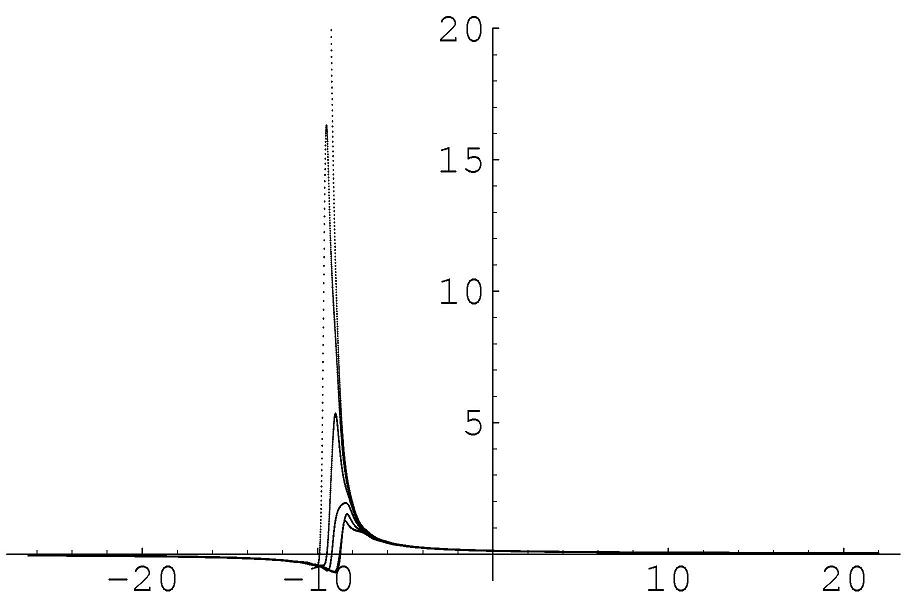}
\caption{\label{fig:af} The coupling constants $g_f$ 
in the case of asymptotic freedom, plotted as a function 
of $\ln{\lambda}/ \ln{b}$ (instead of $s=1/\lambda^2$) 
for 6 values of $f$: $f=0$ (Wegner), $f=0.2$, 0.5, 0.75, 
0.9, and 1. The correspondence between a curve and $f$ 
is such that the curves for larger $f$ reach higher and 
for $f=1$ the corresponding curve apparently shoots 
to infinity around $\lambda \sim |E_{bound state}|$. 
The ultraviolet cutoff is at $b^{16}$, and $b=4$.}
\end{figure}

The behavior of coupling constants in the case of 
asymptotic freedom for various values of $f$ is shown 
in Fig. \ref{fig:af}.
In the $f=1$ case, evolution of $H(s)$ is nearly 
identical to the $f=0$ evolution until the bound 
state begins to emerge. The low-energy plane in 
$\cal V$ moves up to produce the bound-state eigenvalue, 
but the bound state does not decouple from further 
evolution and the low-energy plane simply continues 
to grow apparently indefinitely and our numerical 
calculations fail to converge.

The $f=1$ transformation becomes singular because it 
tries to move the bound-state eigenvalue to the 
lowest-energy diagonal matrix element of the Hamiltonian, as we 
discuss below and explain in greater detail in the 
next section. This means that the bound-state wave 
function is being forced to include only disparately 
small momenta. As a result, the interactions are 
forced to grow to maintain observables at their true 
values. If the evolution is halted before the bound-state 
emerges, the pathological rise of interaction 
terms does not occur yet.

One can see in Fig. \ref{fig:af} not only that the 
larger $f$ the larger the maximal value of the 
coupling constant $g_f$ at the corresponding value 
of the argument $\ln{\lambda_f}/\ln{b}=t_f$, but 
also that {\it simultaneously} the argument $t_f$ 
where the maximum of $g_f$ is reached decreases when $f$ 
increases. In fact, when $f$ increases sufficiently 
to cause a shift of the bound-state eigenvalue by 
one state down the diagonal in comparison to the 
case $f=0$, the maximal coupling constant must 
increase by factor $b$, when $f$ increases sufficiently 
to shift the bound-state eigenvalue by two states down 
the diagonal, the maximal coupling constant must increase 
by factor $b^2$, and so on. Since Fig. \ref{fig:af} 
concerns the case with $b=4$, the maximal values
of the coupling are about 4 and then 16 when the 
bound-state eigenvalue is shifted by one and two 
states down the diagonal, respectively. The next 
shift requires $g_f \sim 64$ and this leads to the 
coupling that grows out of proportion in Fig. 
\ref{fig:af}. When $f=1$, the shift occurs to the 
minimal possible value $t_1=M$, and this requires 
that $g_1$ reaches $b^{|M|} E_{bound state} \sim
b^{25-9} = 4^{16}$ 
in the case illustrated in Fig. \ref{fig:af}. 
With typical machine limitations, numerical 
calculations are expected to fail below the scale 
of binding if one uses the generator $F\{H(s)\}$ 
with $G$ in which $f=1$.

\section{Convergence problems and the $3 \times 3$ matrix}
\label{sec:3x3}

In this section, we explain the phenomenon of rise
of the coupling constant when the SRG parameter
approaches the scale of binding in the
asymptotically free model. After that we will
discuss what happens in the model with limit
cycle. 

We reduce the asymptotically free model to a $3
\times 3$ matrix, allowing $b$ to become arbitrarily
large (i.e., there are only three, strongly coupled
degrees of freedom of drastically different
momentum scales), and show an analytic analysis
that explains the behavior observed in the full
SRG calculation for the $3 \times 3$ matrix. The
$3 \times 3$ matrix model explains the mechanism
that is also at work in large matrices in our
models with asymptotic freedom or limit cycle.

In the $3 \times 3$ case, there is a low-energy
effective coupling that evolves smoothly as the
highest energy scale is decoupled by the SRG
transformation. At a characteristic ``time" $s$,
a $2 \times 2$ low-energy effective Hamiltonian
emerges, characterized by $g_f(s)$, and this
coupling exceeds some critical value when a bound
state emerges. For $f=f_c$, the sign of the
dominant term driving further off-diagonal
evolution changes, and $f_c$ can be computed
analytically in the simple $3 \times 3$ example.
Above $f_c$ the remaining off-diagonal matrix
element is forced to diverge as a power of $b$ to
force the low-energy diagonal to accommodate an
eigenvalue of the wrong magnitude (wrong in the
sense that it is much larger in size than the
corresponding eigenvalue of $T$). The $b
\rightarrow \infty$ limit of the full $3 \times 3$
matrix evolution can be analyzed analytically, and
this is how we explain the mechanism at work in
the full SRG evolution.

Before we analyze the drastically simplified $3
\times 3$ matrix truncation of the asymptotic
freedom model, we recall Wegner's
demonstration~\cite{Wegner:1994} that his
transformation always diagonalizes Hamiltonians,
and we use similar reasoning to explain why a
simpler transformation might fail to converge. 

For any similarity transformation, Tr($H^2$) is 
independent of $s$. Separating diagonal and 
off-diagonal contributions to this trace, one
finds
\beqn
\frac{d}{ds} \sum_m H_{mm}^2 = -\frac{d}{ds} \sum_{m \ne n} \mid H_{mn} \mid^2 \; .
\eeqn
If the magnitudes of diagonal matrix elements 
increase, the magnitudes of off-diagonal matrix 
elements must decrease. Using eq. (\ref{SRGD}) 
for the evolution of the diagonal matrix elements,
\begin{eqnarray}
\frac{d}{ds} \sum_m H_{mm}^2 &=& 4 \sum_{mn} D_m (G_m-G_n) \mid V_{mn} \mid^2 \; \\
&=& 2 \sum_{mn} (D_m-D_n) (G_m-G_n) \mid V_{mn} \mid^2 \; .
\label{diagonals}
\end{eqnarray}
For Wegner's transformation, $G_m =D_m$, so we
have
\beqn
\frac{d}{ds} \sum_m H_{mm}^2 = 2 \sum_{mn} (D_m-D_n)^2 \mid V_{mn} \mid^2 \; .
\eeqn
No term in this sum can be negative, so the only 
way Wegner's transformation can stop driving 
off-diagonal matrix elements to zero is if 
degeneracies appear on the diagonal. In this 
case, the matrix is driven to block diagonal 
form with diagonal degeneracies in any 
non-diagonalized blocks. 

In general, from Eq. (\ref{diagonals}) we see that 
negative terms cannot appear on the right-hand-side 
when all differences $D_m-D_n$ and $G_m-G_n$ always 
satisfy the condition 
\begin{eqnarray}
(G_m-G_n)(D_m-D_n) \ge 0 \, .
\end{eqnarray}
Introducing $(\Delta T)_{mn} = T_m-T_n$ and 
$(\Delta V)_{mn} = V_{mm}-V_{nn}$, one obtains 
for every pair of diagonal elements number $m$ 
and $n$,
\begin{eqnarray}
[f \Delta T + (1 - f )(\Delta T + \Delta V )](\Delta T + \Delta V ) \ge 0 \, ,
\end{eqnarray}
which, by dividing by $\Delta T > 0$ for $m > n$, 
implies that $v = \Delta V/\Delta T$ must satisfy 
the condition
\begin{eqnarray}
[f + (1-f)(1 + v)] ( 1 + v) \ge 0 \, .
\end{eqnarray}
This condition implies for $f \in [0,1]$ that 
either $ v \le \frac{1}{ f-1}$ or $v \ge -1$ and 
only for $f=0$ (Wegner's generator) these two 
regions can join while for $f > 0$ they are 
always disjoint. Instead of differences for 
arbitrary $m$ and $n$, it is sufficient to 
consider differences with $m=n+1$, since all 
differences can be built from the differences 
between neighboring entries on the diagonal. 
Then, in the limit of continuous energy variable, 
the limit of $\Delta T \rightarrow 0$ produces 
the condition 
\begin{eqnarray}
\label{condition}
\frac{ d V}{dT} \le \frac{1}{f-1} \quad \quad  or \quad \quad \frac{d V}{dT} \ge -1 \, , 
\end{eqnarray}
as a sufficient one for the transformation to 
always bring $H$ near the diagonal (outside 
regions of degeneracy mentioned earlier). 

When $f=1$, $G_m=T_m$, and we see that 
convergence can fail if
\beqn
\frac{\Delta V}{\Delta T} < -1
\eeqn
for some momenta. $T_m$ increases monotonically 
with $m$, so we see that problems can appear if 
$V_m$ decreases rapidly with $m$ in some region. 
This is exactly what happens when a negative value 
appears on the diagonal, signaling the appearance 
of a bound-state threshold. The appearance of 
negative values on the diagonal does not guarantee 
that the transformation will stop driving off-diagonal 
matrix elements to zero, and it gives no indication 
that off-diagonal matrix elements will actually start 
to diverge, but it indicates how problems can arise.

To gain further insight, we proceed to a study of 
the $b \rightarrow \infty$ limit, which drastically 
simplifies the couplings between various scales, 
and for further simplicity we truncate the asymptotically 
free Hamiltonian model to a $3 \times 3$ matrix. We refer 
to the three remaining scales as high, middle and low, 
and we write the initial Hamiltonian as:
\beqn
H(0) = \begin{pmatrix}  b & 0 & 0 \\ 0 & 1 & 0 \\ 0 & 0 & \frac{1}{b} \end{pmatrix} \;
-g \begin{pmatrix}  b & \sqrt{b} & 1 \\ \sqrt{b} & 1 & \frac{1}{\sqrt{b}} \\ 1 & \frac{1}{\sqrt{b}} & \frac{1}{b} \end{pmatrix} \;,
\eeqn
where the first matrix is $T$ and the second is 
$V(0)$. We choose $g$ so that the single negative 
eigenvalue is ${\cal O}(1)$ and should appear in 
the middle of the final diagonalized matrix in 
the case $f=0$. These conditions imply that $0.5 < g 
< 1$ and $g$ must be more than ${\cal O}(\frac{1}{b})$ 
away from either extreme. The three eigenvalues are $(1-g)b 
+ {\cal O}(1)$, $(1-2g)/(1-g)+{\cal O}(1/b)$ and 
$\bigl[(1-3g)/(1-2g)\bigr]/b + {\cal O}(1/b^2)$.

We define six couplings in the running interaction,
\beqn
\label{V3x3}
V_f(s) = \begin{pmatrix}  -d_h b & -g_h \sqrt{b} & -g_m \\ 
-g_h \sqrt{b} & -d_m & -\frac{g_l}{\sqrt{b}} \\ 
-g_m & -\frac{g_l}{\sqrt{b}} & -\frac{d_l}{b} \end{pmatrix} \;,
\eeqn
where the diagonal couplings $d_h$, $d_m$ and $d_l$, 
and the off-diagonal couplings $g_h$, $g_m$ and $g_l$ 
are all functions of $s$. At $s=0$ all of these 
couplings are equal to $g$. The scaled interaction 
$\cal V$ is obtained from $V_f(s)$ by replacement 
of $b$ by 1.

Our goal is to obtain accurate estimates for all 
matrix elements of $V_f(s)$, even though their 
magnitudes span many orders. We cannot allow small 
errors in large eigenvalues or far off-diagonal 
matrix elements to produce large errors in small 
diagonal matrix elements that should reproduce 
the eigenvalues of order 1 or $1/b$, the prototypical 
renormalization problem. As $b \rightarrow \infty$ 
this problem can be analyzed analytically.

We find evolution of the coupling constants in 
Eq. (\ref{V3x3}) using Eqs. (\ref{SRGD}) and 
(\ref{SRGV}). The full set of equations for 
arbitrary $f$ is (dots indicate derivatives with 
respect to $s$) 
\begin{eqnarray}
\dot d_h & = &  - 2 b   \, \gamma_{hm} \, g_h^2  
                - 2        \gamma_{hl}  \,g_m^2 \, , 
                \label{eq:couplings1}\\
\dot d_m & = &    2 b^2 \, \gamma_{hm} \, g_h^2  
                - 2 b^{-1} \gamma_{hl}  \,g_l^2 \, , 
                \label{eq:couplings2}\\
\dot d_l & = &    2 b^2 \, \gamma_{hl} \, g_m^2  
                + 2        \gamma_{ml}  \, g_l^2 \, , 
                \label{eq:couplings3}\\ 
\dot g_h & = &  -   b^2 \, \gamma_{hm}[(1-d_h) - b^{-1}(1 - d_m)] \, g_h 
                -         (\gamma_{hl} + b^{-1} \gamma_{ml} ) \, g_m g_l  \, ,
                \label{eq:couplings4}\\
\dot g_m & = &  -   b^2 \, \gamma_{hl}[(1-d_h) - b^{-2}(1 - d_l)] \, g_m  
                -   b   \,(\gamma_{hm} - b^{-1} \gamma_{ml} ) \, g_h g_l  \, ,
                \label{eq:couplings5}\\
\dot g_l & = &      b^2 \,(\gamma_{hm} +        \gamma_{hl} ) \, g_h g_m 
                -          \gamma_{ml}[(1-d_m) - b^{-1}(1 - d_l)] \, g_l \, , 
   \label{eq:couplings6}
\end{eqnarray}
where 
\begin{eqnarray}
\gamma_{hm} & = & \gamma_h - \gamma_m/b  \, , \\
\gamma_{ml} & = & \gamma_m - \gamma_l/b  \, , \\
\gamma_{hl} & = & \gamma_h - \gamma_l/b^2  \, ,\\ 
\gamma_h & = & f + (1-f)( 1- d_h) \, , \\
\gamma_m & = & f + (1-f)( 1- d_m) \, , \\
\gamma_l & = & f + (1-f)( 1- d_l) \, .
\end{eqnarray}

\subsection{Approximate evolution of $3 \times 3$ matrix for $f=0$}

We begin by considering Wegner's transformation, $f=0$. We use Eqs. 
(\ref{eq:couplings1})-(\ref{eq:couplings6}) and keep only the leading terms for large $b$. We will see that all couplings remain ${\cal O}(1)$ when $f=0$, so this analysis is fairly straightforward.
The evolution has two stages: elimination of $g_h$ and $g_m$ in the first stage, and elimination of $g_l$ in the second stage. Terms driving the first stage of evolution are ${\cal O}(b^2)$ and govern evolution until $s$ exceeds ${\cal O}(1/b^2)$. These terms are then exponentially suppressed by $g_h$ and $g_m$, and the second stage of evolution is governed by subleading terms.
The leading terms are:
\begin{eqnarray}
\frac{d}{ds} d_h &=& -2 (1-d_h) g_h^2 b + {\cal O}(1)\;, \label{eq:f0d_h} \\
\frac{d}{ds} d_m &=& 2 (1-d_h) g_h^2 b^2 + {\cal O}(b)\;,  \label{eq:f0d_m} \\
\frac{d}{ds} d_l &=& 2 (1-d_h) g_m^2 b^2 + {\cal O}(1)\;,  \label{eq:f0d_l} \\
\frac{d}{ds} g_h &=& - (1-d_h)^2 g_h b^2 + {\cal O}(b)\;,  \label{eq:f0g_h} \\
\frac{d}{ds} g_m &=& - (1-d_h)^2 g_m b^2 + {\cal O}(b)\;,  \label{eq:f0g_m} \\
\frac{d}{ds} g_l &=& 2 (1-d_h) g_h g_m b^2 + {\cal O}(b)\;. \label{eq:f0g_l}
\end{eqnarray}
We see from (\ref{eq:f0g_h}) and (\ref{eq:f0g_m}) that $g_h=g_m$ during the first stage, because all couplings start at $g$. This implies that $d_m=d_l=g_l$ during this stage also, because the leading equations governing their evolution become identical. Since the low energy $2 \times 2$ submatrix of ${\cal V}$ is determined by $d_m$, $d_l$ and $g_l$, it retains its original form during the first stage of evolution, with a single coupling that can be factored from the submatrix of ${\cal V}$. This is one of the most important results of this analysis and it can be generalized to larger matrices.

The leading term on the right of Eq. (\ref{eq:f0d_h}) is ${\cal O}(b)$ rather than ${\cal O}(b^2)$, so for $s$ of ${\cal O}(1/b^2)$, $d_h$ changes only by ${\cal O}(1/b)$ and we can ignore this change when solving Eqs. (\ref{eq:f0g_h}) and (\ref{eq:f0g_m}), replacing $d_h$ with its initial value, $g$. Solving Eqs. (\ref{eq:f0g_h}) and (\ref{eq:f0g_m}) for large $b$ we obtain
\beqn
g_h(s) \approx g_m(s) \approx g ~{\rm e}^{-(1-g)^2 b^2 s} \;.
\label{eq:f0_ghdecay}
\eeqn
Both of these couplings decay to zero exponentially, and for large $b$ this decay is so rapid that it can be treated as instantaneous. Inserting Eq. (\ref{eq:f0_ghdecay}) in Eq. (\ref{eq:f0d_h}) we find,
\beqn
d_h = g - \frac{g^2}{(1-g) b} +{\cal O}(1/b^2) \;,
\eeqn
at the end of the first stage of evolution.
For a complete leading-order analysis, we need only these leading approximations for $g_h$, $g_m$ and $d_h$. The largest eigenvalue, $(1-d_h) b \approx (1-g) b+{\cal O}(1)$ thus appears in the high-energy corner of the matrix. We do not need the ${\cal O}(1/b)$ correction to $d_h$ to obtain the highest eigenvalue accurately for large $b$ and we will see that we do not need this correction to accurately compute the smaller eigenvalues either.

The leading term governing the early evolution of the remaining $2 \times 2$ submatrix is
\beqn
2 (1-d_h) g_h^2 b^2 \approx 2 (1-g) g^2 b^2 {\rm e}^{-2 (1-g)^2 b^2 s} \approx  \frac{g^2}{1-g} \delta(s) \;,
\eeqn
where $\delta(s)$ is defined so that $\int_0^\infty \delta(s)~ ds = 1$.
The couplings $d_m$, $d_l$ and $g_l$ instantly increase by $g^2/(1-g)$, so at the end of the first stage and the beginning of the second stage of evolution
\beqn
d_m(s) \approx d_l(s) \approx g_l(s) \approx \frac{g}{1-g}+{\cal O}(1/b) \;.
\eeqn
For the range of $g$ that produces a binding energy of ${\cal O}(1)$, these low- and middle-energy couplings exceed 1 and negative values appear on the diagonal of $H$ at the start of the second stage of evolution.

After the initial instant of evolution, $g_h \approx 0$, $g_m \approx 0$ and $d_h \approx g$. Corrections to these approximations have no effect on the leading order of any eigenvalue. For example, the ${\cal O}(1/b)$ correction to $d_h$ has no effect on the smallest eigenvalue, which is ${\cal O}(1/b)$. Returning to the full equations, replacing $g_h$, $g_m$ and $d_h$ with these approximations, choosing an initial value of $g/(1-g)$ for $d_m$, $d_l$ and $g_l$, we find the leading-order equations that govern the second stage of evolution:
\begin{eqnarray}
\frac{d}{ds} d_m &\approx& - 2 (1-d_m) g_l^2 \frac{1}{b} \approx 0 \;, \\
\frac{d}{ds} d_l &\approx& 2 (1-d_m) g_l^2  \;, \\
\frac{d}{ds} g_l &\approx& - (1-d_m)^2 g_l  \;.
\end{eqnarray}

When $g$, the initial coupling, is chosen so that the binding energy is ${\cal O}(1)$, $1-d_m<0$ at the start of the second stage and for large $b$, $d_m$ does not evolve further. This means that the negative eigenvalue, $1-d_m=1-g/(1-g)=(1-2g)/(1-g)$ appears in the middle of the matrix and stays there. This leaves us with,
\begin{eqnarray}
\frac{d}{ds} d_l &\approx& 2 ~\frac{1-2g}{1-g} g_l^2  \;, \\
\frac{d}{ds} g_l &\approx& - \left( \frac{1-2g}{1-g} \right)^2 g_l  \;. 
\end{eqnarray}
From these we see that
\beqn
g_l(s) \approx \frac{g}{1-g}~{\rm e}^{-(1-2g)^2/(1-g)^2~s} \;,
\eeqn
which leads as $s \rightarrow \infty$ to
\beqn
d_l(s) \rightarrow \frac{g}{1-2g} \;.
\eeqn
This means that the smallest eigenvalue, $(1-d_l)/b \approx \bigl[(1-3g)/(1-2g)\bigr]/b$, appears in the infrared corner of the matrix when $f=0$ and no couplings become unnaturally large during the evolution.

\subsection{Approximate evolution of $3 \times 3$ matrix for $f=1$}

Next we study the transformation when $f=1$. 
We again use Eqs. (\ref{eq:couplings1})-(\ref{eq:couplings6}) 
and keep only the leading terms for large $b$ for the first 
stage of the evolution. We will see that $g_h$ and $g_m$ 
again decay exponentially and all other couplings remain 
${\cal O}(1)$ during this first stage, but when $f=1$ the 
coupling $g_l$ grows during the second-stage to 
${\cal O}(b^{1/2})$ before it is finally driven to zero 
and the second-stage's analysis is much more complicated 
because of this unnatural growth. The leading terms for 
$f=1$ are:
\begin{eqnarray}
\frac{d}{ds} d_h &=& -2 g_h^2 b + {\cal O}(1)\;, \\
\frac{d}{ds} d_m &=& 2 g_h^2 b^2 + {\cal O}(b)\;, \\
\frac{d}{ds} d_l &=& 2 g_m^2 b^2 + {\cal O}(1)\;, \\
\frac{d}{ds} g_h &=& - (1-d_h) g_h b^2 + {\cal O}(b)\;, \\
\frac{d}{ds} g_m &=& - (1-d_h) g_m b^2 + {\cal O}(b)\;, \\
\frac{d}{ds} g_l &=& 2 g_h g_m b^2 + {\cal O}(b)\;. 
\end{eqnarray}
These equations govern the first stage of the SRG evolution, with subleading terms becoming important in the second stage. We again see that $g_h=g_m$ to leading order, which implies that $d_m=d_l=g_l$ to leading order during the first stage. The low energy $2 \times 2$ submatrix is determined by these three couplings, and it retains the form of the initial Hamiltonian for all values of $f$ during the first stage of evolution, so we can use the universal coupling governing the submatrix to characterize the transformation for all values of $f$. Solving the equations for $g_h$ and $g_m$ for large $b$, using the fact that $d_h = g + {\cal O}(1/b)$, we obtain
\beqn
g_h(s) \approx g_m(s) \approx g ~{\rm e}^{-(1-g) b^2 s} \;.
\label{eq:f1_ghdecay}
\eeqn
This is nearly the same result we obtained in Eq. (\ref{eq:f0_ghdecay}) for $f=0$ but the exponent is different. Both couplings decay to zero exponentially and $d_h$ changes only at ${\cal O}(1/b)$ because the exponent is ${\cal O}(b^2)$ while the term driving the evolution of $d_h$ is only ${\cal O}(b)$.
The largest eigenvalue, $(1-d_h)b \approx (1-g)b$ again appears in the high-energy corner of the matrix.

The leading term governing the early evolution of the remaining $2 \times 2$ submatrix is
\beqn
2 g_h^2 b^2 \approx 2 g^2 b^2 {\rm e}^{-2 (1-g) b^2 s} \approx \frac{g^2}{1-g} \delta(s) \;,
\eeqn
where $\delta(s)$ is again defined so that $\int_0^\infty \delta(s)~ ds = 1$.
Once again we find that the couplings $d_m$, $d_l$ and $g_l$ instantly increase by $g^2/(1-g)$, to leading order, so at the beginning of the second stage of evolution
\beqn
d_m(s) \approx d_l(s) \approx g_l(s) \approx \frac{g}{1-g} \;.
\eeqn
These second-stage ``initial" values can differ at ${\cal O}(1/b)$, but such differences do not affect the subsequent analysis when $f=0$. We will see that a full analysis when $f=1$ is sensitive to such ${\cal O}(1/b)$ corrections, but we are not interested in the effects of these small corrections. We focus on how the transformation works when the simple scaling analysis breaks down. We want to understand the origin of this breakdown and we can infer its consequences without a derivation of the precise evolution of the couplings. The origin lies in the specific SRG transformation, not in the precise second-stage initial values of the coupling constants.

For the range of $g$ that produces an ${\cal O}(1)$ binding energy, the low- and middle-energy couplings exceed one and negative values again appear on the diagonal of $H$. However, when $f=1$ the appearance of negative values on the diagonal can signal trouble, as argued at the beginning of this section.

After the initial stage of evolution, $g_h \approx 0$, $g_m \approx 0$ and $d_h \approx g$. The subleading terms that govern the second stage of evolution in this case are:
\begin{eqnarray}
\frac{d}{ds} d_m &\approx&  \frac{-2}{b}  g_l^2 \;, \label{eq:dm_f1}\\
\frac{d}{ds} d_l &\approx& 2  g_l^2 \;, \label{eq:dl_f1} \\
\frac{d}{ds} g_l &\approx& - (1-d_m)  g_l  +  \frac{1-d_l}{b} g_l \;. \label{eq:gl_f1}
\end{eqnarray}
When the binding energy is ${\cal O}(1)$, at the start of the second stage $1-d_m<0$ in Eq. (\ref{eq:gl_f1}), so {\it $g_l$ grows exponentially} rather than decaying as it did when $f=0$. Eq. (\ref{eq:dm_f1}) implies that $d_m$ will decrease monotonically at ${\cal O}(1/b)$, but before it can decrease sufficiently to reverse the growth of $g_l$, $g_l$ grows to ${\cal O}(\sqrt{b})$. Meanwhile, Eq. (\ref{eq:dl_f1}) implies that $d_l$ will grow monotonically and it eventually grows to ${\cal O}(b)$ as the bound state eigenvalue is moved from the middle of $H$ to its low-energy corner. Once $d_m$ decreases sufficiently and $d_l$ grows sufficiently, the sign of the right-hand-side of Eq. (\ref{eq:gl_f1}) changes and $g_l$ is driven to zero exponentially from this point.

An exact solution of these equations is readily found because the unitary evolution of the $2 \times 2$ low-energy submatrix of $H$ is simply a rotation that conserves its trace and determinant.
Since $d_m$ decreases monotonically, the bound state eigenvalue cannot appear in the middle of the matrix, as it did when $f=0$. The low-energy eigenvalue that is ${\cal O}(1/b)$ must end up in the middle of the matrix, and this means the precise final result of the transformation is sensitive to ${\cal O}(1/b)$ corrections to the initial value of $d_m$, but the mechanism by which the ${\cal O}(1)$ eigenvalue is transferred to the ${\cal O}(1/b)$ momentum corner of $H$ is not. Thus, the unitarity of the transformation combined with the orders of magnitude of the couplings $d_m$, $g_l$ and $d_l$ at the beginning of the second stage are enough to diagnose why numerical transformations of large matrices fail to converge for some transformations. Moving the bound state eigenvalue to an ``unnatural" location requires growth of couplings by powers of $b$, and in large matrices additional powers of $b$ appear each time the bound state eigenvalue is moved one step down along the diagonal.

\subsection{The critical value $f_c$}

We do not provide complete details of the full evolution for arbitrary $f$ between zero and one; the critical value, $f_c$, is revealed at the end of the first stage of evolution. As the bound-state eigenvalue emerges on the diagonal, $f$ determines whether $g_l$ grows or decays exponentially. 

We again use Eqs. (\ref{eq:couplings1})-(\ref{eq:couplings6}) and keep only the leading terms for large $b$. 
\begin{eqnarray}
\frac{d}{ds} d_h &=& -2 \bigl[1-(1-f) d_h\bigr] g_h^2 b + 
{\cal O}(1)\;, \label{eq:fd_h} \\
\frac{d}{ds} d_m &=& 2 \bigl[1-(1-f) d_h\bigr]
 g_h^2 b^2 + {\cal O}(b)\;,  \label{eq:fd_m} \\
\frac{d}{ds} d_l &=& 2 \bigl[1-(1-f) d_h\bigr] g_m^2 b^2 + 
{\cal O}(1)\;,  \label{eq:fd_l} \\
\frac{d}{ds} g_h &=& - \bigl(1 - d_h\bigr) \bigl[1-(1-f) d_h\bigr] g_h b^2 + 
{\cal O}(b)\;,  \label{eq:fg_h} \\
\frac{d}{ds} g_m &=& -  \bigl(1 - d_h\bigr) \bigl[1-(1-f) d_h\bigr] g_m b^2 +
 {\cal O}(b)\;,  \label{eq:fg_m} \\
\frac{d}{ds} g_l &=& 2 \bigl[1-(1-f) d_h\bigr] g_h g_m b^2 + 
{\cal O}(b)\;. \label{eq:fg_l}
\end{eqnarray}
As expected from the two extremes $f=0$ and $f=1$, during the first stage of evolution, $g_h=g_m$, implying that $d_m=d_l=g_l$. Moreover, the entire analysis of the first stage of evolution is qualitatively independent of $f$. To leading order:
\beqn
g_h(s)=g_m(s)=g~ {\rm e}^{-(1-g)[1-(1-f)g] b^2 s} \;,
\eeqn
{\it cf.} Eqs. (\ref{eq:f0_ghdecay}) and (\ref{eq:f1_ghdecay}).

Once again the integrated effect of these couplings on the low-energy $2 \times 2$ submatrix is to shift the submatrix couplings to a new starting value, at $s=0_+={\cal O}(1/b^2)$,
\beqn
g_l(0_+)=d_m(0_+)=d_l(0_+)=\frac{g}{1-g} \;,
\eeqn
after which we need only examine the equation governing subsequent evolution of $g_l$:
\beqn
\frac{d}{ds} g_l =-\left[1-d_m(s)\right]
\left[1-(1-f) d_m(s)\right] g_l(s)+{\cal O}(1/b) \;.
\eeqn
This off-diagonal coupling grows for $d_m(s) > 1$ if
\beqn
1-(1-f) \frac{g}{1-g} > 0 \;,
\eeqn
from which we determine
\beqn
f_c = 2-\frac{1}{ g} \;.
\eeqn
This value is always between zero and one if there is a bound state with a binding energy that is ${\cal O}(1)$.

The evolution of larger matrices as $b \rightarrow
1$ is far more complicated than the $b \rightarrow
\infty$ limit of the $3 \times 3$ matrix, but
there is remarkable similarity. Any stage in the
evolution of larger matrices can be modeled by a
$3 \times 3$ matrix with the middle diagonal
matched to the point on the diagonal where an
eigenvalue is emerging. We have focused on the
point where a bound-state eigenvalue appears,
because in the evolution of large matrices with
$b$ near 1, it is at this point in the SRG
evolution of the Hamiltonian that transformations
bifurcate; those with $f<f_c$ leave the bound
state eigenvalue on the diagonal where it appears
for $f=0$, while those with $f>f_c$ move it down
the diagonal. Transformations with $f \rightarrow
1$ misplace the bound-state eigenvalue to such
small scales that they fail to numerically
converge. Off-diagonal matrix elements are forced
to diverge exponentially to accomplish this. There
is always a tipping point, $f_c<1$, and at $f_c$
when the bound state emerges, off-diagonal matrix
elements are balanced at an ${\cal O}(1)$ value
between regions of exponential decay and
exponential growth.

\section{Numerical results: limit cycle}
\label{sec:gf}

\begin{figure}[hb]
\includegraphics[scale=0.5]{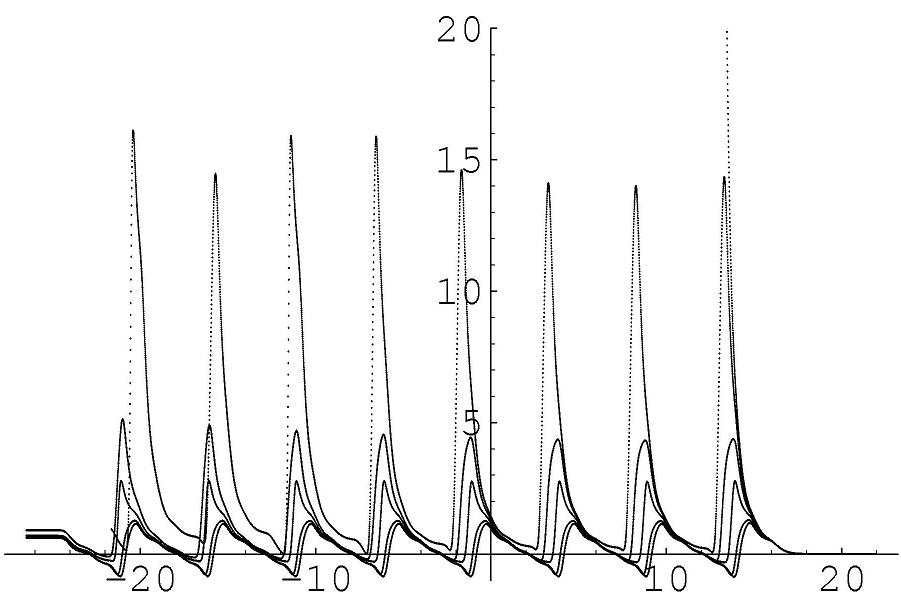}
\caption{\label{fig:lc} The coupling constants $g_f(\lambda)$ in the case 
of a limit cycle, plotted as a function of $\ln{\lambda}/ \ln{b}$ for 6 values
of $f$: $f=0$ (Wegner), $f=0.2$, 0.5, 0.75, 0.9, and 1. The correspondence
between a curve and $f$ is such that the curves for larger $f$ reach higher 
and for $f=1$ the corresponding curve apparently shoots to infinity already 
around $\ln{\lambda}/ \ln{b} \sim 15 $. The ultraviolet cutoff is at $b^{16}$
and $b=4$, see \cite{lcoet}.}
\end{figure}

The limit cycle Hamiltonian of Eq.
(\ref{Hinitial}) with $h \neq 0$ possesses many
bound states, one for each cycle. The number of
eigenvalues that emerge in each cycle is fixed by
$h$. Application of the SRG procedure in the case
of a limit cycle is based on \cite{lcoet}. Numerical
calculation in this case produces Fig.
\ref{fig:lc}. One sees that the case of asymptotic
freedom in the previous subsection corresponds in
its behavior around the scale of binding to one
cycle around the scale of binding in the limit
cycle. In addition, one sees that when $f
\rightarrow 1$ and the SRG generator becomes
purely kinematic, the RG evolution is stuck in the
range of scales corresponding to the scale of
binding in the first cycle (greatest binding
energy). In order to get to the next cycle and
smaller binding energies, one has to introduce
dynamics into the generator $[G,H]$ through the
operator $G$. 

Note that Fig. \ref{fig:lc} also shows that 
the larger the shifts of the bound-state eigenvalues 
down the diagonal and the larger the corresponding 
coupling constant, the more sensitive the numerics 
are to details of the finite matrix, introducing departures 
from a clean cycle. These numerical effects are not fully 
understood but they have no bearing on the findings
reported here that concern the impact of bound states
on the SRG transformations.

It is now clear that in the cases where the
greatest binding energy $E$ is very small, one
does not encounter problems using purely kinematic
$G=T$ as long as $\lambda \gg |E|$. This is our
explanation of what happens in the studies in
nuclear and atomic physics where the binding
energies are much smaller than the cutoffs on
momenta. Hadronic structure is different because
there are no free quarks or gluons; all high
energy eigenstates involve bound states of quarks
and gluons so it is not possible to remain above a
bound-state threshold. 

When one proceeds to the case of a limit cycle with 
a geometric series of binding energies, one must insure that the 
largest of them is small in comparison with desired values of $\lambda$ 
for the generator with $G=T$ to work.

\section{ Conclusion }
\label{sec:c}

The SRG offers an alternative to traditional
renormalization group transformations which
discard degrees of freedom and is being developed
to attack problems where traditional methods fail.
These problems all involve a broad (in principle,
considered infinite) range of momentum scales
that are strongly coupled and are typically not
amenable to perturbation theory, even if the
latter is so-called renormalization-group improved. 
Analytic methods fail and
direct numerical solutions are not possible
because the space of states cannot be truncated,
due to the strong coupling, and the full space is
too large for current or foreseeable numerical
storage and manipulation. In the most interesting
problems one cannot even be sure what operators
are needed to construct renormalizable
Hamiltonians (e.g., what counterterms are required
in the presence of known bare interactions). 

We have not illustrated the initial steps required
to find renormalizable Hamiltonians, choosing
simple model problems in which the required
operators are basically known (e.g., see 
\cite{universality}). But our calculations do
illustrate methods that can make such initial
steps feasible. First, as anticipated in Wilson's
earliest work \cite{Wilson:1965}, we drastically
truncate the space of states by moving from a
continuum to a space of discrete states which are
spaced exponentially, with a tunable spacing
governed by a parameter $b$. Going from a
well-defined continuum problem to a discretized
problem is straightforward in the models we use,
so we have paid little attention to this step; nor
have we shown that with the renormalization
problems we solve the continuum can be
recovered with exponential convergence by letting
$b \rightarrow 1$. The relevance of our
calculations rests on the assumption that such a
limit can be taken with control of errors and
without introducing new renormalization problems.
Once we identify the operators required for
renormalization with large $b$, using
nonperturbative numerical calculations like those
shown here, we reach what is the starting point
for the calculations we present.

Attempts to control the effects of strong coupling
over large numbers of momentum scales typically
fail because divergences emerge. Originally, such
divergences showed up in perturbation theory and
gave rise to the whole renormalization program of
the last century. Far more challenging are the
types of divergence illustrated by our
calculations, divergences that persist if one
attempts to go beyond perturbation theory using
perturbative renormalization schemes such as
standard methods for the renormalization of
quantum electrodynamics. Wilson's renormalization
group improved perturbation theory avoids many of
these problems in asymptotically free theories as
long as cutoffs are kept sufficiently large.
However, to solve problems in which strong-coupling fixed
points \cite{Wilson:1975} or limit cycles exist,
renormalization group transformations often must
be crafted on a case by case basis. What the SRG
offers is a wealth of new transformations and in
this paper we have focussed on some of the
critical features of these new transformations.

The existence of a critical value $f=f_c$ in SRG
transformations with generators $F\{H\}= [G,H]$
and $G = f T + (1-f) D$, is demonstrated in
discrete models with important features such as
asymptotic freedom or a limit cycle. The case of SRG
with $f=1$, or $G=T$, is always in the region $f>f_c$.
Therefore, when bound states exist and the SRG 
parameter $\lambda = 1/\sqrt{s}$ approaches the 
scale of momenta that dominate in the formation 
of bound states, the strength of renormalized 
interactions grows. Numerical calculations of the 
interaction Hamiltonians become increasingly difficult 
due to this growth. But if it is enough 
to calculate effective theories with $\lambda$ 
much larger than the scale of binding, the generator
with $f=1$ can be employed without encountering 
an intractable increase of interaction strength. 

As long as one stays away from the bound-state momentum 
scale, the numerical calculations in the generic 
models are equally powerful for $f > f_c$ as they are 
for $f < f_c$. This result shows that limits on 
applicability of the simplest version of the SRG 
transformation with $f=1$ due to bound states are
not as severe as one might expect provided one keeps
$\lambda$ away from the scale of binding. This is 
important because the SRG transformations with $f=1$ 
are the simplest to implement in the continuum limit 
and in perturbative evaluation of the SRG flow of 
Hamiltonians. 

At the same time, it is also made clear that in order 
to handle cases with a limit cycle, one has to consider 
$f < f_c$. This means that the generator must include
interactions in $G$ and cannot be limited to $T$. The $f=0$
transformation advocated by Wegner \cite{Wegner:1994, 
Kehrein:2006} is able to drive both of our models to diagonal
form. We have not shown that this reproduces both the correct
binding energy and phase shifts for the continuum problem
but such a demonstration should be straightforward and is
left for future work.

Finally, it should be pointed out that this article 
does not resolve many important issues that must
be resolved to deal
with confinement. These go well beyond dealing
with bound states and are not encountered in our
simple generic models, but confinement presents us with
the problem of bound states at all cutoffs and at least in this
respect the limit cycle model should provide important
insights.

\begin{acknowledgments}
We thank  Scott Bogner and Dick Furnstahl 
for useful comments. 
This work was supported in part by the National Science 
Foundation under Grant No.~PHY--0653312.
\end{acknowledgments}

\end{document}